\documentclass[sigconf]{acmart}

\AtBeginDocument{%
  }

\setcopyright{acmlicensed}
\copyrightyear{2018}
\acmYear{2018}
\acmDOI{XXXXXXX.XXXXXXX}

\acmConference[Conference acronym 'XX]{Make sure to enter the correct
  conference title from your rights confirmation emai}{June 03--05,
  2018}{Woodstock, NY}
\acmISBN{978-1-4503-XXXX-X/18/06}

\begin{document}

\title{What's Next? Exploring Utilization, Challenges, and Future Directions of AI-Generated Image Tools in Graphic Design}

\author{Yuying Tang}
\affiliation{%
  \institution{Tsinghua University and Polytechnic University of Milan}
  \city{Beijing and Milan}
  \country{China and Italy}}
\email{tyy21@mails.tsinghua.edu.cn}

\author{Mariana Ciancia}
\affiliation{%
  \institution{Polytechnic University of Milan}
  \city{Milan}
  \country{Italy}}
\email{mariana.ciancia@polimi.it}

\author{Zhigang Wang}
\affiliation{%
  \institution{Tsinghua University}
  \city{Beijing
  }
  \country{China}}
\email{wangzhigang@mail.tsinghua.edu.cn}

\author{Ze Gao}
\authornote{Ze Gao is the corresponding author.}
\affiliation{%
  \institution{Hong Kong University of Science and Technology}
  \city{Hong Kong SAR}
  \country{Hong Kong SAR}}
\email{zgaoap@connect.ust.hk}

\renewcommand{\shortauthors}{Tang et al.}

\begin{abstract}
  Recent advancements in artificial intelligence, such as computer vision and deep learning, have led to the emergence of numerous generative AI platforms, particularly for image generation. However, the application of AI-generated image tools in graphic design has not been extensively explored. This study conducted semi-structured interviews with seven designers of varying experience levels to understand their current usage, challenges, and future functional needs for AI-generated image tools in graphic design. As our findings suggest, AI tools serve as creative partners in design, enhancing human creativity, offering strategic insights, and fostering team collaboration and communication. The findings provide guiding recommendations for the future development of AI-generated image tools, aimed at helping engineers optimize these tools to better meet the needs of graphic designers.

\end{abstract}




\maketitle

\section{Introduction and Related Work}
The recent integration of artificial intelligence (AI) within creativity support tools has significantly expanded the possibilities for human-AI co-creation \cite{sbai2018design, muller2022genaichi, verheijden2023collaborative, guo2023exploring, goodfellow2016deep, vaswani2017attention, brown2020language, elgammal2017can, sun2023ai, tang2023ai, lc2023speculative}. Grounded in Human-Computer Interaction (HCI) theories  \cite{frich2019mapping, li2024understanding, guzdial2019interaction, kantosalo2016modes, muller2011leaving, negrete2014apprentice, rezwana2022understanding, wang2017literature}, extensive research has established frameworks to enhance collaborative creativity between humans and AI.

Researchers have particularly emphasized the necessity of maintaining a balanced human-AI partnership throughout the creative process in the design sector, a key area in creative fields \cite{sbai2018design, wu2021ai, kim2022mixplorer, koch2020imagesense, li2023we, li2023ai, li2024understanding}. Hwang et al. \cite{hwang2022too} investigate how AI-empowered and co-creative tools can address creativity-related challenges in human-AI interactions. In product design, AI facilitates the generation of visuals that, though limited in diversity, foster innovative thinking \cite{lee2023impact, hong2023generative, chiou2023designing}. Gmeiner et al. \cite{gmeiner2023exploring} explore the challenges and opportunities for assisting designers, particularly engineers and architectural designers, in adopting AI-based manufacturing design tools for co-creation. In user interface design, the adoption of diffusion models has led to more intuitive design elements, thus enhancing innovation \cite{cheng2023play, zhang2023layoutdiffusion}. In fashion design, technologies like Attribute-GAN and DreamPose have been pivotal in automating the generation of clothing match pairs and transforming static images into dynamic garment display videos, respectively \cite{liu2019toward, karras2023dreampose}. In graphic design, Engawi \cite{engawi2021impact} delves into AI-driven autonomous branding strategies in Saudi Arabia, assessing their potential challenges and opportunities. Mustafa \cite{mustafa2023impact} examines the impact of AI-assisted and AI-based design tools in graphic design, focusing on website customization, image restoration, branding, logo design, and style transfer.

However, there has been insufficient emphasis on image generation and the influence of designer experience levels on AI tool perceptions, highlighting the need for further research to enhance effective human-AI collaboration in optimizing AI-generated images for graphic design. Meron \cite{meron2022graphic} underscores the importance of bridging the discourse and methodological gap between computer science and graphic design to effectively address issues within AI-assisted design systems. This approach advocates for a collaborative effort between design practitioners and computer scientists to develop AI-supported graphic design systems that are tailored to the needs of graphic designers. Accordingly, our research employs semi-structured interviews with 7 designers of two different experience levels, aiming to understand and address the requirements of graphic designers to optimize AI-generated image tools for graphic design. Our study seeks to answer the following research questions:

\begin{enumerate}
  \item RQ1: How are designers integrating AI-generated image tools into their workflows?
  \item RQ2: How do designers perceive the role of AI-generated images in enhancing versus replacing human creativity in the design process?
  \item RQ3: What capabilities do designers expect from the next generation of AI tools to improve synergistic collaboration?
\end{enumerate}

\section{Methodology}

Our research employed a qualitative approach to address the research questions through semi-structured interviews. We invited seven participants with backgrounds in art and design. The goal of these interviews was to obtain deep insights into the application of AI-generated image tools in graphic design scenarios, specifically focusing on the users' needs and preferences for features within this context.

\subsection{Participants}

Seven art and design background participants were recruited for this study. Although they are not all work as graphic designers, each had practical experience with AI-generated image tools in graphic design. The participants were categorized into two groups based on their experience levels: the first group, consisting of P1, P3, P4, and P7, included designers with three to five years of experience in art and design-related fields, representing the less-experienced level. The second group, comprising P2, P5, and P6, consisted of designers with over five years of experience, representing a more senior level of expertise. The participants aged between 20 and 28 years, with an average age of 24. On average, the participants had been involved in design-related studies or work for about seven years (refer to Table~\ref{tab:participant_data}). The interviews were conducted individually using an online meeting platform to ensure personalized engagement and effective data collection. Each session lasted approximately one hour and was audio-recorded. Transcriptions of these recordings were later analyzed. Informed consent was obtained from all participants for the use of the information recorded during these sessions.

\begin{table}
  \centering
  \caption{Demographic Information of Participants}
  \label{tab:participant_data}
  \begin{tabular}{|p{1cm}|p{1cm}|p{1cm}|p{4cm}|}
    \toprule
    NO.     & Age  & Years in Design & Specific types of studying design/engaging in design work             \\
    \midrule
    P1      & 25.0 & 3.0             & Visual Communication Design                                           \\
    P2      & 26.0 & 8.0             & Product/Service Design                                                \\
    P3      & 20.0 & 3.0             & Product Design                                                        \\
    P4      & 24.0 & 5.0             & Illustration/Visual Communication Design                              \\
    P5      & 24.0 & 10.0            & Illustration/Visual Communication Design, 3D Modeling                 \\
    P6      & 28.0 & 13.0            & Visual Design, Interaction, Digital Strategy and Communication Design \\
    P7      & 23.0 & 5.0             & Graphic Design                                                        \\
    Average & 24.3 & 6.7             &                                                                       \\
    \bottomrule
  \end{tabular}
\end{table}

\subsection{Apparatus and Materials}

The equipment utilized in this study included a laptop and an integrated recording device within the online meeting system. Additionally, we prepared a digital Double Diamond Model diagram for each participant, along with an explanatory video of the Double Diamond Model for those who were not familiar with this concept. The Double Diamond model, frequently employed in the domains of design thinking and the design process \cite{gustafsson2019analysing, pyykko2021approaching, saad2020double}, assists designers in effectively completing projects and promptly addressing design challenges. It serves as an essential method and framework in various design processes. At the beginning of each interview, participants were asked to visualize and illustrate their views on the usage and significance of AI-generated image tools using the Double Diamond model diagram. This approach enabled us to quickly and visually grasp their experiences and personal perspectives. Each interview included 21 open-ended questions designed to understand four key domains: the selection and motivation behind using AI-generated image tools; the integration into and influence on the graphic design process; the ethical, copyright, and professional impacts; and the future expectations and suggestions for improvements in AI-generated image tools for graphic design. The specific interview questions are available in the supplementary materials.

\subsection{Procedure}

The study commenced with an online informed consent process that detailed the background and objectives of the interviews, ensuring that participants were fully aware of the purpose of the study. Subsequently, we gathered background information from the participants (refer to Table~\ref{tab:participant_data}). Next, the interview phase, serving as the core of the study, followed pre-defined question guidelines. To ensure an accurate record of the discussions, we activated a recording device on the online meeting platform, capturing the participants' responses in their entirety to facilitate a comprehensive understanding of their perspectives.

\section{Result}

Our analysis involved examining quotes extracted from 490 minutes of transcribed audio recordings from seven participants. These quotes were utilized to construct a bottom-up hierarchy comprising three high-level themes from nine sub-themes. During the data analysis process, we established connections between these themes and our research questions: Theme 1 explores how designers integrate AI tools into their workflows (RQ1) and assesses the role of AI in augmenting versus replacing human creativity in the design process (RQ2). Theme 2 delves into designers’ perceptions of AI’s role in augmenting versus replacing human creativity (RQ2) and identifies the capabilities designers expect from next-generation AI tools to enhance synergistic collaboration (RQ3). We also discovered that individuals’ perspectives on AI in creative design tend to correlate with their experience, which we have summarized under Theme 3. All participants expressed a preference for using Midjourney for generating AI images, with several specifically highlighting Midjourney’s clearer and easier process.

\subsection{Theme 1: Integration and Influence of AI in the Design Process}
\subsubsection{Application in Design Stages:}
\begin{figure}
  \centering
  \includegraphics[width=0.48\textwidth]{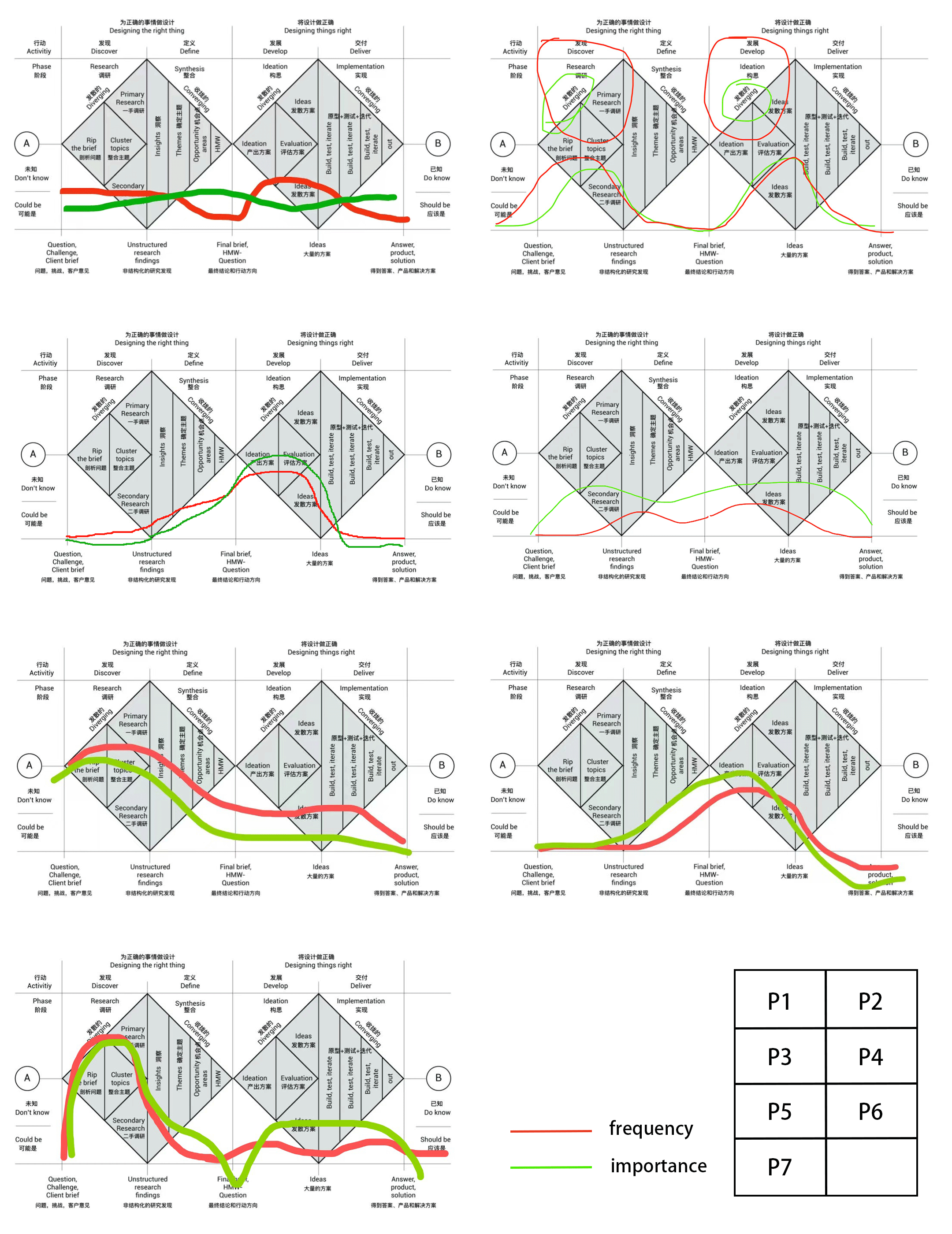}
  \caption{Human-AI Design Process Paintings Based on Double Diamond Model Draw by Seven Interviewees}
  \label{7}
\end{figure}

Participants generally agreed that AI accelerates the creation, inspires, explores new directions, and quickly generates visuals for ideation. They found AI particularly useful in the early inspirational and conceptual design stages. This is reflected in the figures of participants' feedback using the Double Diamond Model, with both the frequency (red wavy lines) and importance (green wavy lines) of AI use showing similar trends in most drawings(Fig.\ref{7}). P3 and P6 rely more on AI in the design process's development stage, mainly used to provide inspiration and reference for visual schemes. P5 and p7 are more inclined to use AI in the design of the Discover stage for scheme research and thinking. In addition, P1, P2, and P7 use AI similarly in the discovery and Development phases. What all respondents have in common is that they rely less on AI in final delivery(Fig.\ref{7}).

\subsubsection{Impact on Creativity and Diversity:}
The impact of AI tools on enhancing creativity and providing diverse ideas was prominently mentioned. P7 likened AI to a virtual library, explaining, ``It is like how you use the library when learning. You cannot take images and results from AI directly, just as you cannot claim a book from a library as your original idea. Humans need to understand they are using a data searcher, like a fast library. You need to remember you're the navigator, the designer. As humans, we need to filter out what we don't need from this library to get the design we want, the information we need, to help push our idea forward... However, without users, the library is not useful." P4 stated, ``AI provides many ideas and improves my creativity," while P5 added, ``AI can provide some results that I haven’t thought of before."

Balancing personal creativity with AI assistance emerged as a critical aspect. Many participants emphasized the importance of human control when using AI image-generation tools, highlighting the crucial role of designers in directing AI through prompt engineering and critically reviewing AI-generated outcomes. This reinforces the ongoing need for human creativity and decision-making in the design process. P2 acknowledged the dual nature of AI's impact, stating, ``AI tools influence our creativity, noting that while AI can enhance productivity, it can sometimes lead to laziness or reliance on AI-generated clichés." However, P6 remarked, ``You also have to think about the right words to give to the algorithm, so I don't know if it's very worth it. And usually, it takes away a lot of time." Despite AI tools augmenting creativity by offering diverse perspectives, concerns persist regarding AI’s limitations in quality consistency, logic-driven design, originality, and contextual understanding, which may affect the depth and uniqueness of creative work. P3 observed, ``AI aids in the creative process, but challenges in using AI tools include limitations in final presentation and the need for adjustments." P5 mentioned, ``AI helps me a lot if I need to do the immediate work. For long-term projects, I will not prefer to use AI, as I don't want to be too influenced by AI, because it can't give me new ideas." In interaction design, P6 noted, ``I think there's not much difference between getting an empty template and an AI-generated sample. So, the user experience you're seeking in your app or project needs to be tailored to the user's needs. Human thoughts are necessary at the beginning and during all the processes; otherwise, it's a bit like making a standard product for everyone. Because the product should follow the user's needs, and needs are different. It's also parallel human thinking."P2 mentioned the need to adapt to AI’s capabilities, which reflects on the balance between AI-assisted creativity and manual design processes. P7 said, ``At the beginning, humans have to control the idea most; you are the person in control. Once you have the ideas, AI can help you to finish your idea. You need to know what you are looking for before using AI." P6 added, ``AI can work in and get more possibilities, so it's not a limitation for me. But the algorithm doesn't know user behaviors or what you want the user to do in your application, even if you explain in detail... that's why it's a bit problematic to create user interfaces with AI, and this is why I don't rely on AI completely for it."

\subsubsection{Improved Efficiency:}
The efficiency of AI tools in managing time constraints was notably emphasized, AI aids in swiftly visualizing a broad range of ideas and styles, proving especially beneficial under tight deadlines. P5 provided a practical example: ``That entire process can be done within five days. And I can also probably do it in three days, which saves me a lot of time".  P7 said: ``To some extent, AI is a super library, but you can find the thing you want easier. AI helps me learn more about creation and possibilities, saving lots of time." P3 mentioned: ``When making products, you can communicate with the team and Party A more quickly and speed up your output and communication in the design process".

\subsubsection{Technical Challenges and Learning Method:}
All the participants noted technical challenges in using AI tools effectively. P3, P4, and P6’s statements underscore this point: ``Challenges in using AI tools include user experience issues and the need for prompt engineering skills... We need to study using AI... Sometimes prompts are similar, but AI will provide different and random results which only include tiny ideas related to my ideas." (P4). ``AI provides me with visual inspiration, but it is time-consuming since I need to delve into the prompts I want to get the right outcomes" (P6). ``natural language is highly required. People should Understand how AI can understand. Most people at the beginning time do not know the high-tech tool. ``The main time spent is to modify the prompt words and select some suitable results to continue to spread and continue to modify in AI" (P3). P5 said: ``System is not perfect. People and AI have different languages. AI still has a long way to learn from us."

The learning method associated with AI tools was also highlighted. All the interviewees said that they learned by watching the experience of using AI tools shared by others on the Internet, as P5 said: ``I need to see some YOUTUBE videos and BiliBili videos, learn how to code and work with Midjourney and ask Midjourney to provide you high results. That’s useful. Each tool has its own method, you need to learn it, even though it is AI", which most designers rely on. However, P3 mentioned that ``the prompts before generating the image are more important, not as important after seeing the generated effect because people can draw it after seeing the effect."

\subsection{Theme 2: Future Expectations and Improvements in AI Tools}
\subsubsection{User-friendly Communication Language and Active Conversational AI:}
Designers struggle with producing work according to their ideas and adjusting prompts, often finding interpreting AI's language and methods time-consuming (P6). They expressed a desire for AI tools that better understand complex prompts, provide more detailed and realistic outputs, and offer constructive feedback, thereby improving the human-AI collaborative process and design works. Improvements in communication would enable easier interaction and more precise translation of creative ideas into AI-generated outcomes,  transforming AI into a digital partner rather than just a tool. As P2 said: ``ChatGPT is a simple interface good example, being popular".

\subsubsection{Customization and Style Consistency:}
\begin{figure}[H]
  \centering
  \includegraphics[width=0.48\textwidth]{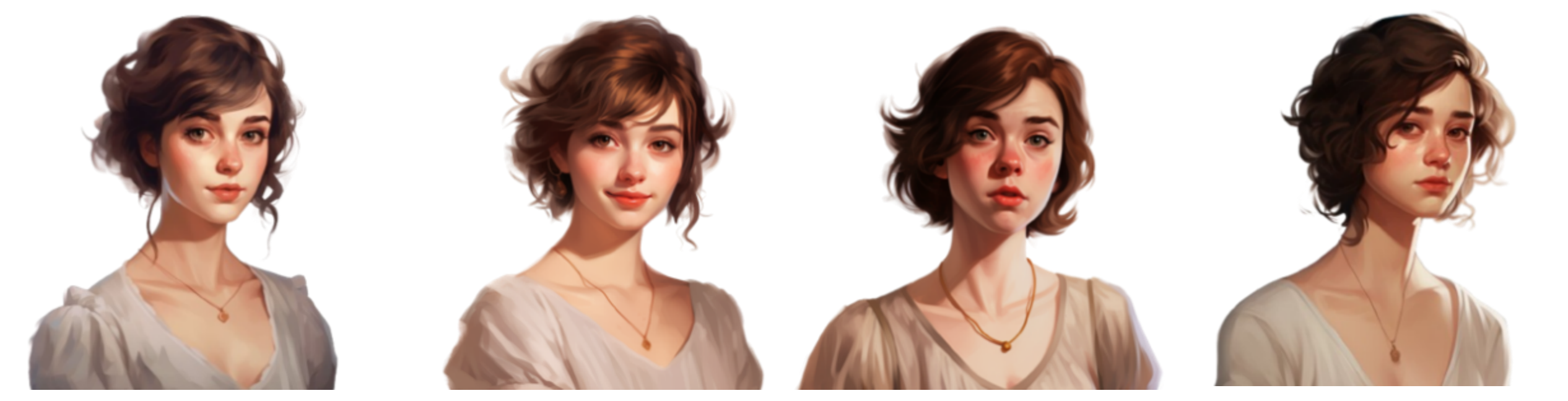}
  \caption{A Series of Graphics Provided and Generated by P3}
  \label{series}
\end{figure}

There is a strong need for AI tools capable of adapting to individual content styles and maintaining consistency across outputs. P2 and P6 highlight the need for AI to cater to diverse user groups with specific features and functions. AI struggles with final production and output, often leading to inconsistent quality and errors(Fig.\ref{series}). Many participants highlighted the challenge of unifying styles in a series of graphics (P3, P4). Therefore, the ability to customize AI responses and ensure consistent aesthetics is crucial for professional use.

\subsection{Theme 3: The Impact of Different Experience Levels on Designers' View of AI Tools}

\subsubsection{AI-Driven Design Process}

Designers with less-experienced levels, such as P1, P3, P4, and P7, tend to use AI tools less frequently. Therefore, when discussing specific instances, they provide detailed descriptions of singular cases. For instance, P1 talks about one rural IP image design, P3 discusses a board game design, P4 focuses on one layout design assignment, and P7 elaborates on creating a magazine for a local school.

In contrast, P2, P5, and P6, who have a more advanced understanding of AI in design with senior experience levels, shared how AI aids them in diverse scenarios. And they have their own specific workflow to complete design tasks with AI. P2, for example, has developed a personalized AI design workflow. In his process, AI usage is segmented into two distinct phases. The initial phase involves conducting 10 to 20 experiments with varying prompts to articulate ideas, predominantly using text-to-image to derive ideas from basic concepts. Subsequently, P2 refines the prompts based on previous outcomes and embarks on the second phase, which mainly relies on image-to-image techniques. This phase uses sketches or new elements and images from the first phase, integrating different styles and colors inspired by the initial round. P5 underscores the need for uniqueness in their AI-generated images, striving to ensure these images are distinct from existing ones, based on their personal knowledge, memories, and opinions. P5 also mentioned that ``what aspects I care about always be defined by the aims and application scenarios I need to apply for. " They also stress the importance of achieving optimal results within a limited timeframe. As P6 said: ``you need to give them a file where they can work because I think you use them for a purpose. If you use them purposelessly, then you can create anything you want, but a designer should like to have a useful outcome for, for example, me, I gave rules, I gave information, and then I tried to generate content." It can be seen that less experienced designers still stay in individual projects to use AI to assist a specific task of design, while experienced designers have their own evaluation criteria and methods for AI, and have formed their AI design method framework.

\subsubsection{Key Points of Concern in AI Image}

Different levels of designers exhibit distinct focal points when using AI to generate images. P1, P3, P4, and P7, designers with less-experienced levels, focus on specific aspects of AI-generated images. For instance, P1 emphasizes the importance of logical consistency and appealing colors, P3 mentions logic and function, P4 focuses on color and human character depiction, and P7 is concerned about the extent of detail control in the images.

In contrast,  P2, P5, and P6, more seasoned designers with an advanced understanding of AI in design, prioritize the overall impact of the images. P2 looks for an overall feel that accurately reflects their ideas, without a specific emphasis on color. P5 expresses a desire for AI to expedite the creative process in an overview version instead of using hand drawing. P6 emphasizes the rendering effect, encompassing a comprehensive overview.

This variation highlights the different levels of needs and expectations designers have for AI-generated images based on their experience. Experienced designers, who possess a wealth of hands-on skills, tend to view AI-generated images as a source of overall inspiration and a reference for their work. They prefer to use their creativity to design specific details rather than relying on AI. In contrast, less experienced designers place higher demands on the details of AI-generated images. This reflects their greater reliance on AI due to a possible lack of their own methodologies and hands-on abilities in design tasks.

\subsubsection{Impact on the Design Profession}

Experienced designers (P2, P5, P6) express the belief that professional designers will not be completely supplanted by AI. In contrast, junior designers (P1, P3, P4, P7) exhibit a more pessimistic outlook. Due to their uncertainty about future trends and a lack of professional experience, they speculate that AI might eventually supersede human designers.

P3 noted, "Designers in the mid-level bracket could be impacted by AI, potentially leading to job losses." P4 elaborated on the notion of workforce reduction, suggesting that AI could serve as an alternative to human employees. P7 opined, "If you have more creativity, you may not lose your jobs."

P2 contends, "Professionals in the field are unlikely to lose their positions, whereas entry-level designers might be more susceptible to replacement." P5 observed, "AI could offer new opportunities for companies less concerned with the intricacies of design. Such businesses might opt for AI solutions over employing designers. Consequently, designers may transition to roles centered around AI operation and management. This shift represents an evolution in working methods due to AI advancements. Designers are unlikely to face obsolescence; instead, they should focus on acquiring new skills, such as proficiency in AI interfacing." P6 argued, "It's implausible to envisage a scenario where designers are entirely replaced by AI, as it marks a revolution rather than a replacement. AI cannot replicate human creativity and thought processes."




\section{Discussion}

The interviews provided insights into the current utilization and anticipated future of AI-generated image tools in graphic design. These tools are primarily used as assistants during the initial stages of the design process, where they inspire creativity through a variety of ideas and visualizations. However, their effectiveness is limited by the need for human oversight, ethical considerations, and the limitations of current AI technology. Participants expressed concerns about the ethical use and copyright implications of AI-generated images. P5 highlighted these issues, stating, "Copyright issues and the ethical use of AI-generated images," which underscore the complexities involved in determining ownership and originality of AI-generated content. Looking ahead, designers envision AI in design becoming more sophisticated, intuitive, and synergistically integrated with human creativity, enhancing rather than replacing human skills. The findings specifically reflect two main aspects as follows:

\subsection{New models for AI assisting rather than replacing designers}

\subsubsection{Balancing Human Creativity with AI in Design Processes}
Leveraging the potential of generative AI necessitates focusing on human values, aiming to enhance – not replace – the skills, voices, and rights of human creators. Human involvement remains a pivotal component of the design and creative process, regardless of whether artists feed AI with rudimentary keywords or a story outline, blend different AI tools, or designers instruct AI to generate moving images based on a given static one. As yet, no fully autonomous AI-generated images exist. Designers highlight the necessity of maintaining control and guiding AI through effective, prompt engineering, as exemplified by Participant P1's practice of critically reviewing AI-generated results. This approach emphasizes the vital role of human judgment, recognizing that AI, while helpful in providing visual inspiration and rapid iteration, lacks the nuanced understanding inherent to human designers. Consequently, designers ensure AI outputs align with design projects' specific backgrounds, goals, and audiences. Beyond these, the potential of AI to usurp human creativity sparks contentious debates. While it is conceivable that AI might mimic certain creative processes, it lacks the reservoir of lived experiences, emotional nuance, and intuitive grasp of the human condition that frequently undergird truly transformative creativity. Consequently, it is improbable that AI will entirely supplant human creativity.

\subsubsection{Lead Designers Focus on Creative Tasks}
AI's capacity to handle repetitive and less creative tasks was appreciated. Participant P5 discussed her approach to handling various commercial projects, noting the importance of discerning specific client needs and desired outcomes to determine the extent of personal creative input versus AI utilization. P5 observed that many clients do not fully understand what constitutes good design and often prefer conventional approaches. In such cases, excessive creativity might lead to dissatisfaction, and AI-generated results could align more closely with these clients' expectations. However, for clients with a high appreciation for art, aesthetics, and design thinking, P5 would invest significant personal creativity and minimize AI assistance in the creative process. This allows designers to focus more on creative aspects of design, leveraging AI for basic and low-level creative tasks such as essential layout generation, color scheme testing, and initial mock-ups.

\subsubsection{Support for Diverse Creative Processes}
AI tools' capacity to provide unexpected visual inspirations was valued for expanding designers' horizons and enhancing their ability to undertake various design tasks, enhancing their knowledge and skills across different design domains. ``Without AI humans have more learning processes need more time and learn more things (P2)." For instance, an illustrator like Participant P5 can gain inspiration for transforming illustrations into poster designs and obtain layout and image ideas from AI tools, moreover, for those like Participant P5 and P3. Without 3D modeling skills at the beginning, AI aids in the early stages of design by quickly providing a 3D preview of their ideas (P5). ``I am mainly from an industrial design background, but I need to do graphic design occasionally, and I am not good at drawing illustrations. When I try to make a storyboard, I draw it hard. Still, the AI-generated will quickly convey what I want to express" (P3).

\subsection{Desired Features and Capabilities for AI Design Tools}

\subsubsection{AI's Partnerships in Detailed and Direct Feedback and Guidelines}
Designers expressed a strong need for AI tools capable of generating content and providing constructive feedback, essential in learning and creativity. Participants P1 and P5 shared their experiences of often being uncertain about how AI perceives human design concepts, how it evaluates the generated content, and how to modify prompts for optimization. They described a time-consuming learning process through online tutorials and community participation, where they studied how others crafted more effective prompts. This process, while informative, often fails to address specific issues they encounter. P1 envisions a direct brainwave-to-image output, suggesting a future where thinking creatively becomes a human edge. Therefore, there is a growing desire for future AI to offer faster and more direct advice and guidance, particularly in areas such as evaluating creative inspiration and refining design prompts. P6 also suggests real-time collaboration and sharing features in AI tools, similar to platforms like Figma. Designers increasingly envision AI tools evolving into creative partners or digital tutors, providing suggestions for improvement and opening new avenues for creativity.

\subsubsection{The Future of Artificial Intelligence Tools in Team Dynamics}
In team collaborations, AI tools are emerging as pivotal in visualizing ideas, fostering brainstorming and discussions, and accelerating the creative process. These tools introduce diverse and rapid collaborative approaches. P3 highlighted the role of AI in enhancing communication within teams, noting, ``AI aids in clearly drawing designs essential for effective communication between designers, technical members, and business members. This clarity is vital when simple descriptions are insufficient, and better drawings are required, which AI can proficiently generate." P4 focused on the application of AI in idea generation and research, stating, ``AI tools are currently well-suited for collaborative design work, aiding in brainstorming and idea generation." They also noted the diversity AI brings to the creative process, ``I use Midjourney for generating images and ideas, influencing collaboration within design teams by combining ideas from team members with varied backgrounds." Many interviewees expressed the desire for AI to play a more significant role in team collaboration in the future. AI is seen as promising in facilitating communication, bridging skill gaps among team members, and enabling more efficient collaboration within design teams. P7 echoed this sentiment, describing how AI ``facilitates communication and collaboration within design teams," underscoring the growing importance of AI in enhancing team dynamics and interdisciplinary collaboration. However, there are challenges in fully integrating AI into team collaboration. P2 pointed out current limitations, ``As of now, it's hard to use AI in collaboration. We can only share chats with AI among friends. There are only a few tools that allow each team member to interact with their AI, but other team members can't jointly access the same AI." This highlights an area for future development in AI tools to support further and streamline team collaboration.


\section{Conclusion and Future Work}

This research highlights designers' aspirations for AI to serve as a collaborative force that enhances rather than replaces human creativity. It is essential to maintain human oversight and align AI objectives with designers' needs to ensure ethical and synergistic co-creation. The findings suggest that designers envision future AI as a conversational partner that understands design principles and adapts to creative styles. The aim is to achieve human-AI synergy by combining complementary strengths—AI's efficiency and computational power with human judgment, emotions, and intuition. Designers seek AI that continuously learns and tailors its capabilities to both the task and the individual, moving beyond functioning as rigid tools. Specifically, future developments in AI-generated image tools should focus on advancing natural language processing for intuitive communication, integrating these tools into graphic design processes, and promoting transparency, trust, and complementarity between human designers and AI systems. Additionally, conducting research on human-AI creativity and fostering interdisciplinary collaboration is vital to realizing this vision. By embracing this cooperative approach, AI can become an invaluable partner, amplifying originality and expanding the boundaries of human imagination.

However, as preliminary research, our study has its limitations: With only seven interviewees, the qualitative insights, while valuable, may not fully capture the spectrum of opinions and experiences among designers. Conducting more interviews could provide a richer, more nuanced understanding of individual perspectives. Furthermore, this study represents a snapshot in time. Long-term tracking or follow-up studies could reveal how designers' needs and AI tool usage evolve as technology advances.

\bibliographystyle{ACM-Reference-Format}
\bibliography{main}
\end{document}